\documentclass[aps,prd,onecolumn,groupedaddress,nofootinbib,amssymb,nopacs
]{revtex4}
\usepackage[dvips]{graphicx}
\usepackage{amssymb}
\usepackage{amsmath}
\usepackage{graphicx,color}
\usepackage{amsfonts}
\usepackage{bm}
\usepackage{cancel}
\usepackage{comment}
\allowdisplaybreaks[4]
 
\usepackage{hyperref}

\begin{document}

\tolerance=5000

\title{Late time attractors of some varying Chaplygin gas cosmological models}

\author{
M. Khurshudyan$^{1,2,3}$\footnote{E-mail: khurshudyan@ice.csic.es, khurshudyan@yandex.ru}, 
R. Myrzakulov$^{4,5}$\footnote{Email: rmyrzakulov@gmail.com}
}
\affiliation{
$^{1}${\small {\em Institute of Physics, University of Silesia, 40-007 Katowice, Poland}}\\ 
$^{2}${\small {\em Consejo Superior de Investigaciones Cient\'{\i}ficas, ICE/CSIC-IEEC, Campus UAB, Carrer de Can Magrans s/n, 08193 Bellaterra (Barcelona) Spain}}\\
$^{3}${\small {\em International Laboratory for Theoretical Cosmology, Tomsk State University of Control Systems and Radioelectronics (TUSUR), 634050 Tomsk, Russia}}\\
$^{4}${\small {\em Eurasian National University, Nur-Sultan 010008, Kazakhstan}}\\
$^{5}${\small {\em Ratbay Myrzakulov Eurasian International Centre for Theoretical Physics, Nur-Sultan 010009, Kazakhstan}}\\
}

\begin{abstract}

The goal of this paper is to study new cosmological models where the dark energy is a varying Chaplygin gas. This specific dark energy model with non-linear EoS had been often discussed in modern cosmology. Contrary to previous studies, we consider new forms of non-linear non-gravitational interaction between dark matter and assumed dark energy models. We applied the phase space analysis allowing understanding the late time behavior of the models. It allows demonstrating that considered non-gravitational interactions can solve the cosmological coincidence problem. On the other hand, we applied Bayesian Machine Learning technique to learn the constraints on the free parameters. In this way, we gained a better understanding of the models providing a hint which of them can be ruled out. Moreover, the learning based on the simulated expansion rate data shows that the models cannot solve the $H_{0}$ tension problem.

\end{abstract}

\maketitle

\section{Introduction}\label{sec:INT}

In modern cosmology, there are several key open problems and various approaches to solve them including gravitational particle creation and modification of general relativity\cite{Liddle:1010476, Yoo:2012ug, Clifton:2011jh, Bousso:2012dk, Villata:2013ata, Velten:2014nra, Bamba:2012cp, Overduin:2001pv, Tian:2020tur, Khurshudyan:2017yjc, Li:2019loh, Elizalde:2018dvw, Cai:2019bdh, Aljaf:2020eqh} (and references therein). Recently another one known as the $H_{0}$ tension problem has been added to this list \cite{Aghanim:2018eyx, Riess:2018byc, Wong:2019kwg, Freedman:2019jwv}. The goal of this paper is (1) to consider various new cosmological models explaining the late time accelerated expansion of the universe~\cite{Riess:1998cb, Perlmutter:1998np, Spergel:2003cb, Tegmark:2003ud, Abazajian:2004it, Abazajian:2004aja, Hawkins:2002sg, Verde:2001sf}, and (2) to see whether or not the models solve the $H_{0}$ tension problem\cite{Amirhashchi:2020qep, Sharov:2020bnk, Yang:2021flj, Braglia:2020iik, Yao:2020pji, Elizalde:2020pps, Kreisch:2019yzn, Alestas:2020mvb, Elizalde:2020mfs, DiValentino:2020naf, Elizalde:2021kmo}. The analysis of the models is based on two approaches. In particular, we use phase space analysis in the first part of the paper, while in the second part of the paper we use Bayesian Machine Learning to learn the constraints on the model parameters (see for instance~\cite{Elizalde:2020pps, Elizalde:2020mfs, Elizalde:2021kmo, Aljaf:2020nsl}). The phase space portrait of a model is one of the mechanisms providing a qualitative understanding of the model. It allows to observe all states of the model without solving a system of differential equations, but solving algebraic equations \cite{Chen:2008ft, Jamil:2012yz, Leon:2012mt, Shabani:2013djy, Escobar:2011cz, Jarv:2004uk, Xu:2012jf, Leon:2012vt, Yang:2010vv}  to mention a few. On the other hand, Bayesian Machine Learning based on the generative process allows one to infer crucial properties of the model directly from the model used in the generative process. This method was recently applied in several studies indicating very interesting departures from traditional approaches used in cosmology. We will come back to this in the second part of the paper when we discuss learned constraints on the parameters.

In general, dark fluids are actively used to explain the accelerated expansion of the universe. Chaplygin gas is one of such dark fluids \cite{Gorini:2004by, Kamenshchik:2001cp, Bento:2002ps, Sandvik:2002jz}
\begin{equation}\label{eq:CH}
P_{de} = A\rho_{de} - \frac{B}{\rho_{de}^{\alpha}},
\end{equation}
where $A$, $B$ and $\alpha$ are positive constants to be constrained from observational data and $\rho_{de}$ it is the energy density of the gas (or dark energy fluid). In literature there are various modifications of this fluid too \cite{Naji:2014dya, Guo:2005qy, Chimento:1995da, UrenaLopez:2000aj, Khurshudyan:2014ewa, Kahya:2015dpa, Fu:2009zza} (to mention a few). 

In this paper, we will consider one of them, assuming that the parameter $B$ in Equation (\ref{eq:CH}) is not constant. In general, a varying Chaplygin gas can be constructed assuming both $A$ and $B$ parameters in Equation (\ref{eq:CH}) are not constant. However, in each case, it is important to have a modification either based on some very well motivated physics or obtain the constraints in order to justify the crafted phenomenological modification. In our work, we follow to the second approach and having phenomenological modifications we will learn which one can survive. It should be mentioned that the interest towards Chaplygin gas is its dark energy and dark matter unifying ability. However, some critics on this issue also exist (see for instance \cite{Sandvik:2002jz}).

In our analysis we consider FRW universe with ($c=8\pi G = 1$)
\begin{equation}\label{eq:Fridmmanvlambda}
H^{2}=\frac{\dot{a}^{2}}{a^{2}}=\frac{\rho}{3},
\end{equation}
\begin{equation}\label{eq:fridman}
\frac{\ddot{a}}{a}=-\frac{1}{6}(\rho+3P),
\end{equation}
where $\rho$ it is the energy density of the effective fluid, while $P$ it is the pressure. It is easy to see from the structure of Equations (\ref{eq:Fridmmanvlambda}) and  (\ref{eq:fridman}), that, for example, an assumption about the effective fluid will allow closing the system of differential equations. In general, we can assume that the effective fluid can be multicomponent one that needs to be confirmed the observational data. However, here we follow a simplified scenario assuming that the energy source entering in Equations (\ref{eq:Fridmmanvlambda}) and (\ref{eq:fridman})  is two-component one and that
\begin{equation}\label{eq:rhoeff}
\rho = \rho_{dm} + \rho_{de},
\end{equation}
and 
\begin{equation}\label{eq:Peff}
P  = P_{dm} + P_{de},
\end{equation}
where $\rho_{de}$ and $P_{de}$ are the energy density and pressure of dark energy, while $\rho_{dm}$ and $P_{dm}$ are the energy density and pressure of dark matter. Now the question is how to be with the dark energy and dark matter. A phenomenological assumption about the content of the universe is a commonly accepted approach in modern cosmology, which will be used in this paper as well. There are various assumptions about the nature and origin of dark energy and each of them had been investigated from various perspectives. On going research towards dark fluid representation of dark energy, for instance, allows to develop class of viscous dark fluids equally applicable to cosmic inflation and the accelerated expansion of the late time universe \cite{Avelino:2013wea} (and references therein). Moreover, it has been shown that dark fluids can be described by linear and non-linear EoS. On the other hand, it appears that dark fluids can be solutions of algebraic and differential equations \cite{Nojiri:2005sr}. Besides dark fluids we can use, for instance, scalar field to construct viable dark energy models \cite{Khurshudyan:2014yva} (and references therein). Quintessence and phantom dark energy models are among them. However, our not complete understanding of the universe shows that dark energy independent of its way of interpretation can be not enough. In this regard, various studies allowed to introduce a phenomenological idea of interacting dark energy models demonstrating that it can be equivalently good as the other ideas. Practically there is nothing against to this idea. An interaction between dark components of the universe can smooth some unpleasant aspects in the dynamics of the models making them more attractive. The main mechanism to implement a non-gravitational interaction is based on the energy transfer between dark components in the following way
\begin{equation}\label{eq:firstfluid}
\dot{\rho}_{dm}+3H(\rho_{dm}+P_{dm})=Q,
\end{equation}
and
\begin{equation}\label{eq:secondfluid}
\dot{\rho}_{de}+3H(\rho_{de}+P_{de})=-Q,
\end{equation}
where $Q$ is the notation of non-gravitational interaction\cite{Yang:2007pa, Cai:2009ht, Wei:2010cs, Guo:2004xx, Wei:2006tn, Cai:2004dk, Zhang:2005jj, Wu:2007zzf, Li:2008uv,Chen:2008ft, He:2009pd, Chimento:2009hj, Setare:2006wh, Setare:2006sv, Setare:2007bx,Jamil:2009zzb, Baldi:2010vv, He:2008tn, Khurshudyan:2013oba, Hakobyan:2013hca, Sadeghi:2013aga, Sadeghi:2013xca, Xu:2013iw, Ma:2009uw, Arevalo:2011hh, Khurshudyan:2015mva, Aljaf:2019ilr, Aljaf:2020eqh, Sadri:2019yqs, Elizalde:2018ahd, Odintsov:2017icc} (Non-gravitational interaction introduced in Equations (\ref{eq:firstfluid}) and (\ref{eq:secondfluid}) indicates energy transfer between dark matter and dark energy). There are different phenomenological parameterizations of the interaction term and some of them can help to solve/alleviate the $H_{0}$ tension problem too (see references of this paper for more details about interacting dark energy models). 

Up to this point we discussed our goal and shortly mentioned the tools we will use. Then, we presented how to describe interacting dark energy models and what is the background dynamics. But yet we did not discuss the varying Chaplygin models we will consider. We will consider two particular models of varying Chaplygin gas (see for instance \cite{Naji:2014dya} and \cite{Fu:2009zza})
\begin{equation}\label{eq:M1}
P_{de} = A\rho_{de} - \frac{BH^{-n}}{\rho_{de}^{\alpha}}, 
\end{equation}
and
\begin{equation}\label{eq:M2}
P_{de} = A\rho_{de} - \frac{Ba^{-n}}{\rho_{de}^{\alpha}},
\end{equation}
where $H$ is the Hubble parameter, $a$ is the scale factor, while $n$ is a constant, will be used to be the dark energy. Eventually, we see from Equations (\ref{eq:firstfluid})  and (\ref{eq:secondfluid}) that another crucial aspect allowing us to study the cosmological models is the interaction term Q. In other words, it should be given in order to close the system of differential equations describing the background dynamics of the universe with interacting dark energy models. In this paper, we consider non-linear interactions Q to be presented in Section \ref{sec:IntModels}. It should be mentioned, that we assume dark matter is a pressureless fluid.

To end this section, let us mention that the phase space analysis allows immediately demonstrate, that considered forms of non-gravitational interactions allow solving the cosmological coincidence problem. Besides the phase space analysis we applied Bayesian Machine Learning approach and learned the constraints on the parameters of each model. It gives a hint that within considered models the $H_{0}$
tension cannot be solved. Moreover, we found which of the models eventually needs to be ruled out.

The paper is organised as follows: In Section \ref{sec:IntModels} , we will discuss how the phase space analysis can be implemented to find late time attractor solutions for suggested new cosmological models. In Section \ref{sec:PSA}, the phase space analysis is performed, late time attractors are found and classified according to their cosmological applicability. Moreover, in Section \ref{sec:FFS}, for each model the constraints on the parameters using Bayesian Machine Learning have been learned. In the same section very briefly the crucial aspects of the approach have been discussed too. Finally, discussion on obtained results are summarised in Section \ref{sec:Discussion}.

\section{Interacting models and autonomous system}\label{sec:IntModels}

In the literature there is a huge number of works devoted to the phase space analysis of various cosmological models. In order to start the phase space analysis of a model, an appropriate autonomous system should be found first which is a system of algebraic equations to be solved. The critical points are solutions of the autonomous system. They are stable if appropriate Jacobian matrix has a negative trace and positive determinant. This is in case of linear stability. Following \cite{Xu:2013iw} for our models we set

\begin{equation}\label{eq:x}
x = \frac{\rho_{de}}{3H^{2}},
\end{equation}
\begin{equation}\label{eq:y}
y = \frac{P_{de}}{3H^{2}},
\end{equation}
\begin{equation}\label{eq:z}
z = \frac{\rho_{dm}}{3H^{2}},
\end{equation}
and 
\begin{equation}\label{eq:N}
N=\ln{a},
\end{equation}
where $a$ it is the scale factor. It is not hard to see, that for interacting models the autonomous system reads as
\begin{equation}\label{eq:xPrime}
x^{\prime} = 3x (1+y) - \frac{Q+3H\rho_{de} (1+\omega_{de})}{3H^{3}},
\end{equation}
and
\begin{equation}\label{eq:yPrime}
y^{\prime} = 3y(1+y) + \frac{\dot{P}_{de}}{3H^{3}},
\end{equation}
where $\prime$ is the derivative with respect to $N$, dot is the derivative with respect to cosmic time, $Q$ is interaction term. Explicit forms of Equations (\ref{eq:xPrime}) and (\ref{eq:yPrime}) are obtained when the form of $Q$ is given. It is easy to see, that physically reasonable solutions should satisfy $0 \leq x \leq 1$~($0 \leq x_{de} \leq 1$) and $0 \leq z \leq 1$~($0 \leq z_{de} \leq 1$) constraints. On the other hand, a stable critical point will be an attractor, which we are looking for. At the same time we should remember that $x$ and $z$ according to Equations (\ref{eq:Fridmmanvlambda}), (\ref{eq:x}) and (\ref{eq:z}) should satisfy to the following constraint
\begin{equation}\label{eq:FConst}
x+z =1.
\end{equation}
Moreover, after some algebra we can see that the EoS parameter of the varying Chaplygin gas in terms of $x$ and $y$ reads as
\begin{equation}\label{eq:EoSC}
\omega_{de} = \frac{y}{x},
\end{equation}
while the EoS parameter of the effective fluid reads as
\begin{equation}\label{eq:EoSeff}
\omega_{eff} = \frac{P_{de}}{\rho_{de} + \rho_{dm}} = y.
\end{equation}
In the above equation, we used the fact that dark matter is cold and pressureless. On the other hand, it is not hard to show that the deceleration parameter $q$ reads as
\begin{equation}\label{eq:q}
q = -1 - \frac{\dot{H}}{H^{2}} = \frac{1}{2}(1 + 3 y).
\end{equation}
Now, in order to obtain the explicit forms of Equations (\ref{eq:xPrime}) and (\ref{eq:yPrime})  we need to define the form for $Q$. As we mentioned earlier we will consider new non-linear non-gravitational interactions. What we consider here can be obtained from the following more general form of interaction $Q$
\begin{equation}\label{eq:Int}
Q = 3H b \left( \rho + \frac{\hat{\rho}}{\rho}\right ),
\end{equation}
where $\rho$ could be either the energy density of the effective fluid, or the energy density one of the components of the effective fluid. $\hat{\rho}$ it is product of the energy densities of the components of the effective fluid i.e. three possibilities could be $\rho_{de}^{2}$, $\rho_{dm}^{2}$ and $\rho_{de}\rho_{dm}$. In this way we can note that the equations discussed in this section are self consistent and allow performing the phase space analysis. In the next section we will discuss our results.

\section{Phase space analysis}\label{sec:PSA}

As we mentioned earlier, in order to have late time attractors for the cosmological models, we need the explicit form of non-gravitational interaction $Q$ to have Equations (\ref{eq:xPrime}) and (\ref{eq:yPrime}) determined. In this work we will pay attention to fixed sign non-linear interactions following from Equation (\ref{eq:Int}). Non-linear interactions having similar structure as here, including also models with non-linear sign-changeable interactions already have been considered in \cite{Khurshudyan:2015mva}. At this stage of the analysis to simplify the discussion, we impose some constraints on the model parameter taken from the literature. In particular, we will impose
\begin{equation}\label{eq:alpha}
0 < \alpha \leq 1,
\end{equation}
\begin{equation}\label{eq:n}
0 \leq n \leq 5,
\end{equation}
constraints. On the other hand, the constraints on the deceleration parameter $q$
\begin{equation}
-1 \leq q < 0,
\end{equation}
\begin{equation}
0 < r = \frac{\Omega_{m}}{\Omega_{de}} \leq 1
\end{equation}
and on the EoS parameter of the varying Chaplygin gas
\begin{equation}\label{eq:omegaconst}
-2 \leq \omega_{de} < 0,
\end{equation}
gave us an option to reduce the phase space size allowing to find late time scaling attractors more efficiently. On the other hand, it is not excluded that the phantom divide about $z\approx 0.2$
 redshift took place in our universe, therefore constraint Equation (\ref{eq:omegaconst}) is considered \cite{Nesseris:2006er, Hu:2004kh, MohseniSadjadi:2009va}.. Having imposed discussed constraints we deal with conditional attractors. They allow obtaining constraints depended late time states of the universe. The goal of this section is to find and discuss stable critical points preparing appropriate initial seeds allowing to initialize Bayesian Machine Learning-based analysis of the models.

\subsection{Varying Chaplygin gas $P_{de} = A\rho_{de} - \frac{BH^{-n}}{\rho_{de}^{\alpha}}$}\label{ssec:CG1}

We start the study from the models where the varying Chaplygin gas, Equation (\ref{eq:M1}), interacts with cold dark matter and the the forms of non-linear interactions are defined from Equation (\ref{eq:Int}).
 
\subsubsection{Interaction $Q = 3 H b \left( \rho_{de} + \frac{\rho_{dm}^{2}}{ \rho_{de} + \rho_{dm}} \right )$}\label{ssec:Q1}

The first cosmological model we study is a model where the interaction between the varying Chaplygin gas, Equation (\ref{eq:M1}), and cold dark matter is given as

\begin{equation}\label{eq:Int1}
Q = 3 H b \left( \rho_{de} + \frac{\rho_{dm}^{2}}{ \rho_{de} + \rho_{dm}} \right )
\end{equation}
and real critical points are presented in Table~\ref{tab:table1}. 
\begin{table}
  \centering
    \begin{tabular}{ | c  | c  | c  | c  | c | c |  p{5cm} |}
    \hline
 $S. P.$ & $x$ & $y$ & Type of stability\\
      \hline
 $E.1.1$  & $\frac{b-1+ \sqrt{1+(2-3b)b}}{2b}$ & $-1$ & stable node\\
          \hline
$E.1.2$  &  $\frac{A-b + \sqrt{(A-b)(A+3b)} }{2 (A-b) }$ & $ \frac{A(A-b + \sqrt{(A-b)(A+3b)}) }{2(A-b)}$ & unstable node\\
    \hline
  $E.1.3$  &  $\frac{A-b - \sqrt{(A-b)(A+3b)} }{2 (A-b) }$ & $ \frac{A(A-b - \sqrt{(A-b)(A+3b)}) }{2(A-b)}$ & stable focus\\
    \hline
    \end{tabular}
\caption{Critical points corresponding to interacting varying Chaplygin gas, Equation (\ref{eq:M1}), for the non-gravitational interaction $Q$ given by Equation (\ref{eq:Int1}).}
  \label{tab:table1}
\end{table}

In this case, $3$ different physically reasonable critical points exist and only two of them are stable~($E.1.1$ and $E.1.3$). However only the critical point $E.1.1$ due to imposed constraints is a late time scaling attractor with
\begin{equation}
r = \frac{\Omega_{dm}}{\Omega_{de}} = \frac{1-b - \sqrt{1+ (2-3b) b}}{2(b-1)},
\end{equation} 
tending to a constant. For this model the decelerated parameter is $q=-1$ and $\omega_{eff} = -1$, while the EoS parameter of the varying Chaplygin gas Equation~(\ref{eq:M1}) reads as
\begin{equation}
\omega_{de} = - \frac {2b} { b - 1 +  \sqrt {1+ (2-3b) b} }
\end{equation}

It is not hard to see that considering varying Chaplygin gas, Equation (\ref{eq:M1}), is a phantom dark energy. Discussed behavior is obtained when the parameters satisfy to Equations (\ref{eq:alpha}) and (\ref{eq:A}) and $0 < b < 2/3$. The critical point $E.1.3$ will be stable when $0 < \alpha \leq 1$, $0 \leq n \leq 5$, $b=0$ and $A > 1+ \alpha + n/2$. This solution describes a matter dominate state of the universe, since $x=0$, and in such universe the accelerated expansion is not possible. Therefore, only $E.1.1$ will be the physically reasonable solution where the universe will reach starting from the state described by $E.1.2$.

\subsubsection{Interaction  $Q = 3 H b \left( \rho_{de} + \frac{\rho_{dm} \rho_{de}}{ \rho_{de} + \rho_{dm}} \right )$}\label{sseq:Q2}

In the second model the interaction between varying Chaplygin, gas Equation (\ref{eq:M1}), and cold dark matter is taken to be
\begin{equation}\label{eq:Int2}
Q = 3 H b \left( \rho_{de} + \frac{\rho_{dm} \rho_{de}}{ \rho_{de} + \rho_{dm}} \right ).
\end{equation}

In this case, physically reasonable two critical points exist and they are presented in Table \ref{tab:table2}. The study shows that $E.2.1$ is a physically reasonable solution, moreover, it is a late time scaling attractor when $0 \leq n \leq 5$, $0 < \alpha \leq 1$, $A \geq 0$ and $0 < b < 2/3$. This solution represents a state of the universe where varying Chaplygin gas, Equation (\ref{eq:M1}), is a phantom dark energy with
\begin{equation}
\omega_{c} = -\frac{1+2b + \sqrt{1+4b^{2}} }{2}
\end{equation} 
This model is free from the cosmological coincidence problem due to
\begin{equation}
r = \frac{\Omega_{dm}}{\Omega_{de}} = \frac{2b -1 + \sqrt{1+ 4b^{2}}}{2}.
\end{equation}

\begin{figure}[h!]
 \begin{center}$
 \begin{array}{cccc}
 \includegraphics[width=80 mm]{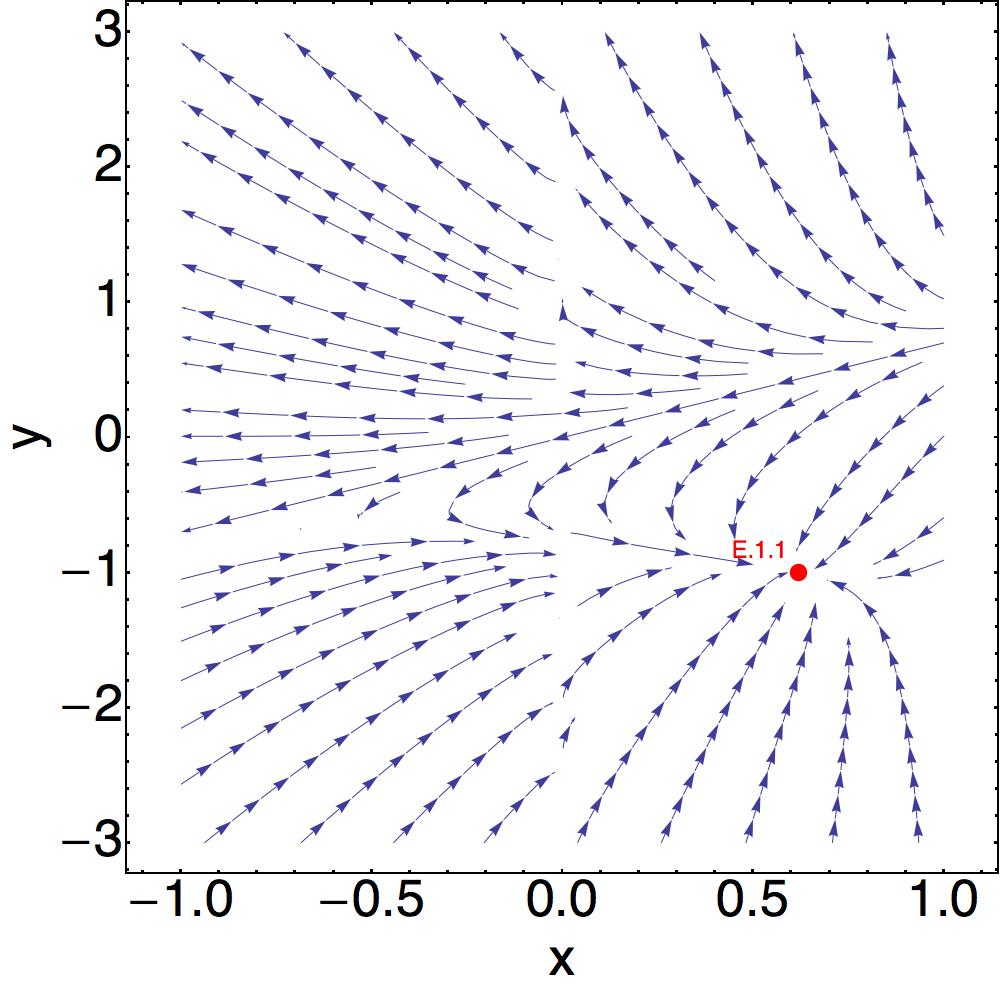}  &
 \includegraphics[width=80 mm]{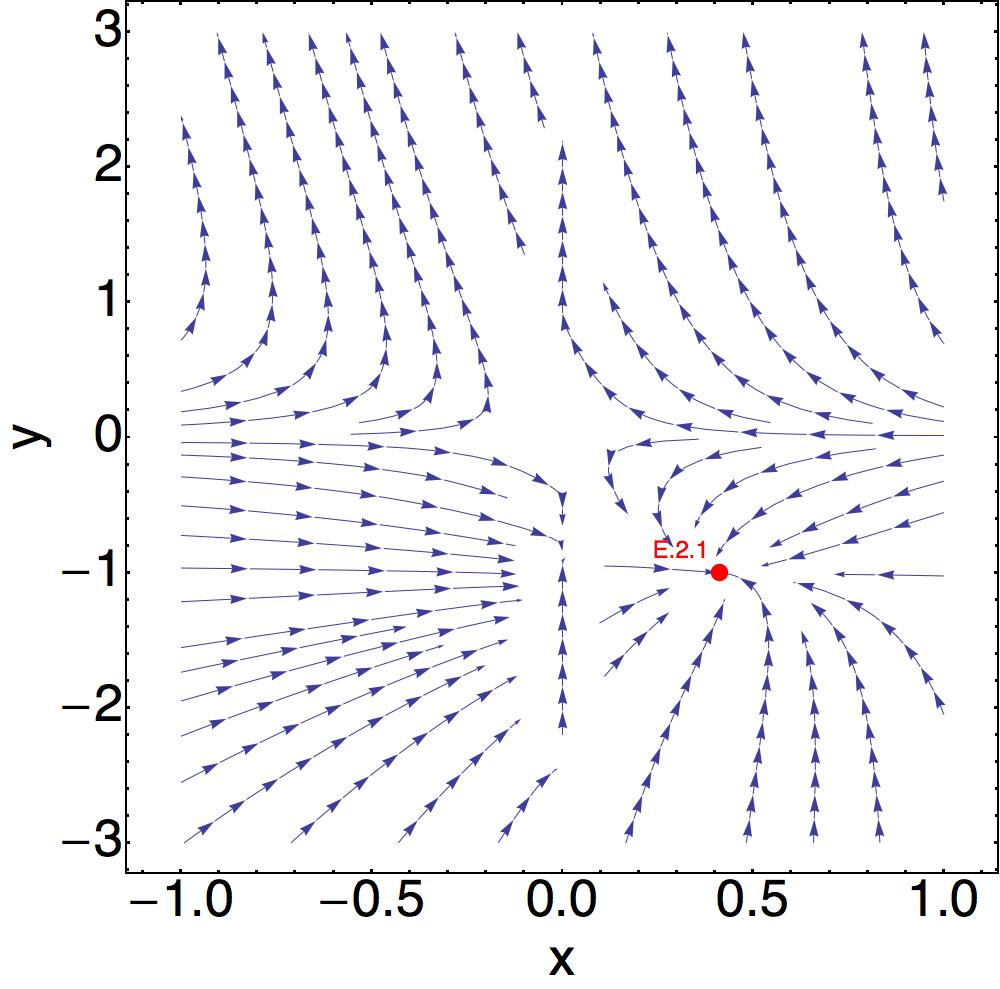}  \\
 \end{array}$
 \end{center}
\caption{Phase space portraits for interacting varying Chaplygin gas, Equation (\ref{eq:M1}). The left plot represents the model where the non-gravitational interaction $Q$ is given by Equation (\ref{eq:Int1}). The right plot represents the model where the non-gravitational interaction $Q$ is given by Equation (\ref{eq:Int2}).}
 \label{fig:Fig1}
\end{figure}

Critical point $E.2.2$ is physically reasonable solution when $b=0$ i.e. $x =1$ and $y=A$ (when $A > 0$), and this is a state of the universe where the accelerated expansion is not possible. Moreover, the varying Chaplygin gas, Equation (\ref{eq:M1}), is a usual fluid and completely dominates the dynamics of the universe. As was expected $E.2.2$ is not a stable critical point. Figure (\ref{fig:Fig1}) represents the phase space portraits for $E.1.2$  and $E.2.1$ critical points. For a symmetry in plots, we considered $-1 \leq x \leq 1$ interval, however, we should remember that  $x \in (0,1]$.

\begin{table}
  \centering
    \begin{tabular}{ | c |  c |  c | c |  c |  c |  p{3cm} |}
      \hline
 $S. P.$ & $x$ & $y$ & Type of stability\\
      \hline
$E.2.1$  &  $\frac{1+2b - \sqrt{1+ 4b^{2}} }{2b}$ & $-1$ & stable node\\
    \hline
 $E.2.2$  & $\frac{A+2b}{A+b}$ & $\frac{A(A+2b)}{A+b}$ & unstable node\\

    \hline
               \end{tabular}
\caption{Critical points corresponding to interacting varying Chaplygin gas, Equation (\ref{eq:M1}), for the non-gravitational interaction $Q$ given by Equation (\ref{eq:Int2}).}
  \label{tab:table2}
\end{table}

\subsubsection{Interaction $Q = 3 H b \left( \rho_{de} +\rho_{dm} + \frac{\rho_{dm}^{2}}{ \rho_{de} + \rho_{dm}} \right )$} \label{ssec:Q3}

Another cosmological model has been studied in this paper assuming the interaction term $Q$ has the following form

\begin{equation}\label{eq:Int3}
Q = 3 H b \left( \rho_{de} +\rho_{dm} + \frac{\rho_{dm}^{2}}{ \rho_{de} + \rho_{dm}} \right ).
\end{equation}
In this case, $3$ critical points are physically acceptable (among $4$ critical points) presented in Table \ref{tab:table3}. The critical point $E.3.1$ is a late time scaling solution since
\begin{equation}
r = \frac{\Omega_{dm}}{\Omega_{de}} = \frac{\sqrt{1- 4 b^{2}}}{2(1-2b)} - \frac{1}{2},
\end{equation}
and represents the state of the universe, where the varying Chaplygin gas Equation (\ref{eq:M1})  is phantom dark energy with
\begin{equation}
\omega_{de} = -\frac{2b}{2b - 1 + \sqrt{1-4b^{2}}}.
\end{equation}

This is a late time attractor when $0 \leq n \leq 5$, $0 < b \leq 2/5$ and $0 < \alpha \leq 1$. On the other hand, the critical points $E.3.2$ and $E.3.3$ are physically reasonable when $b = 0$ and the critical points are unstable. The phase portrait indicates that in considered case the evolution of the universe started from one of the unstable states (either $E.3.2$ or $E.3.3$) eventually will evolve to the state described by $E.3.1$ solution. It is not hard to see that in case of $E.3.3$ the universe will be in a matter dominating state and the accelerated expansion is not possible. In considered $3$ models of this subsection we observed, that appropriate late time scaling attractors describe the state of the universe with phantom varying Chaplygin gas Equation (\ref{eq:M1}). Moreover, the solution of the cosmological coincidence problem exists. Therefore, it is important to learn the constraints on the free parameters to have better understanding of the models. To this we will come in the second part of this work.
 
\begin{table}
  \centering
    \begin{tabular}{ | c | c | c | c | c | c | c | p{5cm} |}
    \hline
 $S. P.$ & $x$ & $y$ & Type of stability \\
      \hline
 $E.3.1$  & $\frac{2b - 1 + \sqrt{1-4b^{2}}}{2b}$ & $-1$ & stable node \\
          \hline
$E.3.2$  &  $\frac{A-2b + \sqrt{A^{2} + 4Ab - 4b^{2}} }{2(A-b)}$ & $ \frac{A(A-2b + \sqrt{A^{2} + 4Ab - 4b^{2}}) }{2(A-b)}$ & unstable node\\
    \hline
  $E.3.3$  &  $\frac{A-2b - \sqrt{A^{2} + 4Ab - 4b^{2}} }{2(A-b)}$ & $\frac{A(A-2b - \sqrt{A^{2} + 4Ab - 4b^{2}} )}{2(A-b)}$ & unstable node or unstable focus \\
    \hline
    
    \end{tabular}
\caption{Critical points corresponding to interacting varying Chaplygin gas, Equation (\ref{eq:M1}), for the non-gravitational interaction $Q$ given by Equation (\ref{eq:Int3}).}
  \label{tab:table3}
\end{table}

\subsection{Varying Chaplygin gas $P_{de} = A\rho_{de} - \frac{Ba^{-n}}{\rho_{de}^{\alpha}}$}

In Section \ref{ssec:Q1}, Section \ref{sseq:Q2} and Section \ref{ssec:Q3} we considered cosmological models where varying Chaplygin gas was given by Equation (\ref{eq:M1}). In the next three subsections we will consider cosmological models, where varying Chaplygin gas is given by Equation (\ref{eq:M2}).

\subsubsection{Interaction $Q = 3 H b \left( \rho_{de} + \frac{\rho_{dm}^{2}}{ \rho_{de} + \rho_{dm}} \right )$} \label{ssec:Q4}

The model with the varying Chaplygin gas, Equation (\ref{eq:M2}), when non-gravitational interaction is given by Equation ({\ref{eq:Int1}}) has $4$ critical points (Table \ref{tab:table4}, where $E.4.3$ is not physically reasonable solution). The study shows that the considered model has only one conditional late time attractor corresponding to $E.4.4$ solution. Moreover, the study shows that for imposed constraints on the parameters of the model we will obtain the states of the universe with a quintessence varying Chaplygin gas, Equation (\ref{eq:M2}). However, for some combinations of the values of the free parameters (from initial constraints imposed on the parameters) $E.4.4$ solution can describe the state of the universe where the varying Chaplygin gas, Equation (\ref{eq:M2}), is a phantom. Observed dual possibility shows an interesting departure of this model from previously considered models. Moreover, we found that the late time attractor $E.4.4$ is scaling attractor. In particular, to simplify the analysis, let us demonstrate that $E.4.4$ is a scaling attractor when $0 \leq b \leq 1/10$ and $n=1$. It is easy to see, that in this case, the attractor $E.4.4$ is scaling because
\begin{equation}\label{eq:rE44}
r = \frac{\Omega_{dm}}{\Omega_{de}} = \frac{1}{2} \left ( -1 + \frac{2 + 3 \alpha + 9 b (1+\alpha)}{ \sqrt{ (2+3\alpha - 3b (1+\alpha)) (2+3\alpha + 9 b(1+\alpha))} } \right ).
\end{equation}

Moreover, considered constraints on the parameters provide the deceleration parameter $q$
\begin{equation}
q = -1 + \frac{1}{2(1+\alpha)},
\end{equation}
to be $0 < q \leq -1$. On the other hand, we can see that the varying Chaplygin gas, Equation (\ref{eq:M2}), is a quintessence dark energy with 
\begin{equation}
\omega_{de} = - \frac{2b (2+3\alpha)}{3b (1+\alpha) - 2 - 3\alpha + \sqrt{(2 - 3b - 3\alpha (b-1)) (2 + 3\alpha + 9b (1+\alpha))} }.
\end{equation}

The solution of the cosmological coincidence problem in form Equation (\ref{eq:rE44}) is different from the solutions presented in Section \ref{ssec:CG1} and clearly demonstrates that the solution can be obtained due to the non-gravitational interaction. In summary, starting the evolution from one of the initial states described by $E.4.1$ and $E.4.2$, the universe eventually will end up on the state described by $E.4.4$.

\begin{table}
  \centering
    \begin{tabular}{ | c |  c |  c |  c |  c  |c |  c | p{5cm} |}
    \hline
 $S. P.$ & $x$ & $y$ & Type of stability\\
      \hline
 $E.4.1$  & $\frac{A-b + \sqrt{(A-b)(A+3b)}}{2(A-b)}$ & $\frac{A(A-b + \sqrt{(A-b)(A+3b)})}{2(A-b)}$ & unstable node\\
          \hline
$E.4.2$  &  $\frac{A-b - \sqrt{(A-b)(A+3b)}}{2(A-b)}$ &  $\frac{A(A-b - \sqrt{(A-b)(A+3b)})}{2(A-b)}$ & unstable node\\
    \hline
 $E.4.3$  &  $\frac{n+3(b-1)(1+\alpha)-\hat{r}}{6b(1+\alpha)}$ & $-1+\frac{n}{3(1+\alpha)}$ & --\\
    \hline
 $E.4.4$  &  $\frac{n+3(b-1)(1+\alpha)+\hat{r}}{6b(1+\alpha)}$ & $-1+\frac{n}{3(1+\alpha)}$ & stable node or stable focus\\
    \hline
    \end{tabular}
\caption{Critical points corresponding to interacting varying Chaplygin gas, Equation~(\ref{eq:M2}), for the non-gravitational interaction $Q$ given by Equation~(\ref{eq:Int1}). $\hat{r}=\sqrt{(n- 3(1+\alpha)(1+3b)) (n - 3(1+\alpha)(1-b))}$.}
 \label{tab:table4}
\end{table}

\subsubsection{Interaction  $Q = 3 H b \left( \rho_{de} + \frac{\rho_{dm} \rho_{de}}{ \rho_{de} + \rho_{dm}} \right )$}\label{sseq:Q5}

The three critical points obtained for the cosmological model with the varying Chaplygin gas, Equation (\ref{eq:M2}), and non-gravitational interaction $Q$, Equation (\ref{eq:Int2}), are presented in Table \ref{tab:table5}.

 \begin{table}
  \centering
    \begin{tabular}{ | c | c |  c |  c |  c |  c |  c |  c | p{5cm} |}
    \hline
 $S. P.$ & $x$ & $y$ & Type of stability \\
      \hline
 $E.5.1$  & $\frac{A+2b}{A+b}$ & $\frac{A(A+2b)}{A+b}$ & unstable node\\
          \hline 
$E.5.2$  &  $\frac{3 (1+2b) (1+\alpha)- n - \hat{r}}{6b(1+\alpha)}$ &  $- 1 + \frac{n}{3(1+\alpha)}$ & stable node \\
    \hline
$E.5.3$  &  $\frac{3 (1+2b) (1+\alpha)- n + \hat{r}}{6b(1+\alpha)}$ &  $- 1 + \frac{n}{3(1+\alpha)}$ & --\\
    \hline
     \end{tabular}
\caption{Critical points corresponding to interacting varying Chaplygin gas, Equation~(\ref{eq:M2}), for the non-gravitational interaction $Q$ given by Equation~(\ref{eq:Int2}), where $\hat{r} = \sqrt{3b b^{2}(1+\alpha)^{2} + (n-3(1+\alpha))^{2}}$.}
  \label{tab:table5}
\end{table}

The study shows that the late time attractor $E.5.2$ can represent the states of the universe where the varying Chaplygin gas, Equation (\ref{eq:M2}), either is a quintessence dark energy (Table \ref{tab:table6}) or a a phantom dark energy (Table \ref{tab:table7}) depends on the values of the parameters of the model (the general constraints on the parameters are the same as in previous cases).
 
\begin{table}
  \centering
    \begin{tabular}{  | c | c |  c |  c |  c |  c |  c |  c | p{2cm} |}
    \hline
 $n$ & $\alpha$ & $b$ \\
 
      \hline
 $0<n\leq \frac{3}{2}$  & $0\leq \alpha \leq 1$ & $0<b\leq \frac{n}{3 \alpha +n+3}$ \\
 
          \hline
$\frac{3}{2}<n<2$  &  $0\leq \alpha \leq \frac{1}{3} (2 n-3)$ &  $ 0<b\leq \frac{6 \alpha -2 n+6}{9 \alpha +9}$ \\

    \hline
$\frac{3}{2}<n<2$  &  $\frac{1}{3} (2 n-3)<\alpha \leq 1$ &  $ 0<b\leq \frac{n}{3 \alpha +n+3}$ \\
    \hline
    
    $2\leq n\leq 3$  &  $\frac{n-2}{2}<\alpha \leq \frac{1}{3} (2 n-3)$ &  $0<b\leq \frac{6 \alpha -2 n+6}{9 \alpha +9}$ \\
    \hline
    
   $2\leq n\leq 3$   &  $\frac{1}{3} (2 n-3)<\alpha \leq 1$ &  $0<b\leq \frac{n}{3 \alpha +n+3}$ \\
    \hline
    
    $3<n<4$  &  $\frac{n-2}{2}<\alpha \leq 1$ &  $0<b\leq \frac{6 \alpha -2 n+6}{9 \alpha +9}$ \\
    \hline
     \end{tabular}
\caption{Constraints on the model parameters for the interacting varying Chaplygin gas, Equation~(\ref{eq:M2}), with non-gravitational interaction term $Q$ given by Equation~(\ref{eq:Int2}). $E.5.2$ describes the state of the universe where the varying Chaplygin gas is a quintessence dark energy.}
  \label{tab:table6}
\end{table}
 
 \begin{table}
  \centering
    \begin{tabular}{ | c | c |  c |  c |  c |  c |  c |  c | p{2cm} |}
    \hline
 $n$ & $\alpha$ & $b$ \\
 
      \hline
 $n=0$  & $0\leq \alpha \leq 1$ & $0<b<\frac{6 \alpha +6}{9 \alpha +9}$\\
 
          \hline
$0<n<\frac{3}{2}$  &  $0\leq \alpha \leq 1$ &  $\frac{n}{3 \alpha +n+3}<b\leq \frac{6 \alpha -2 n+6}{9 \alpha +9}$ \\

    \hline
$\frac{3}{2}\leq n<3$  &  $ \frac{1}{3} (2 n-3)<\alpha \leq 1$ &  $\frac{n}{3 \alpha +n+3}<b\leq \frac{6 \alpha -2 n+6}{9 \alpha +9}$ \\
    \hline

     \end{tabular}
\caption{Constraints on the model parameters for the interacting varying Chaplygin gas, Equation~(\ref{eq:M2}), with non-gravitational interaction term $Q$ given by Equation~(\ref{eq:Int2}). $E.5.2$ describes the state of the universe where the varying Chaplygin gas, Equation~(\ref{eq:M2}), is a phantom dark energy.}
  \label{tab:table7}
\end{table}

The phase space portrait of this models shows that for imposed constraints, the evaluation of the universe will start from the state described by $E.5.1$ (unstable node) and will reach to the state one described by $E.5.2$ (stable node). On the other hand, $E.5.3$ is physically not reasonable solution, while $E.5.2$ is a scaling attractor
\begin{equation}
r = \frac{3(1-2b)(1+\alpha) - n - \hat{r}}{2(n - 3(1+\alpha))},
\end{equation} 
with

\begin{equation}
q = -1 + \frac{n}{2(1+\alpha)},
\end{equation}

\begin{equation}
\omega_{eff} = -1 + \frac{n}{3(1+\alpha)},
\end{equation}

\begin{equation}
\omega_{de} = - \frac{3(1+2b)(1+\alpha) - n + \hat{r}}{6(1+\alpha)},
\end{equation}
where  $\hat{r} = \sqrt{3b b^{2}(1+\alpha)^{2} + (n-3(1+\alpha))^{2}}$.

\subsubsection{Interaction $Q = 3 H b \left( \rho_{de} +\rho_{dm} + \frac{\rho_{dm}^{2}}{ \rho_{de} + \rho_{dm}} \right )$}\label{sseq:Q6}

\begin{table}
  \centering
    \begin{tabular}{ | c | c |  c |  c |  c |  c |  c |  c | p{5cm} |}
    \hline
 $S. P.$ & $x$ & $y$ & Type of stability \\
      \hline
 $E.6.1$  & $\frac{A-2b + \sqrt{A^{2} + 4Ab - 4b^{2}}}{2(A-b)}$ & $\frac{A(A-2b + \sqrt{A^{2} + 4Ab - 4b^{2}})}{2(A-b)}$ & unstable node \\
          \hline 
$E.6.2$  &  $\frac{A-2b - \sqrt{A^{2} + 4Ab - 4b^{2}}}{2(A-b)}$ & $\frac{A(A-2b + \sqrt{A^{2} + 4Ab - 4b^{2}})}{2(A-b)}$ & stable focus\\
    \hline
$E.6.3$  &  $\frac{n + 3 (- 1+2b) (1+\alpha)  - \hat{r}}{6b(1+\alpha)}$ &  $- 1 + \frac{n}{3(1+\alpha)}$ & -\\
    \hline
$E.6.4$  &  $\frac{n + 3 (- 1+2b) (1+\alpha) + \hat{r}}{6b(1+\alpha)}$ &  $- 1 + \frac{n}{3(1+\alpha)}$ & stable node or stable focus\\
    \hline
     \end{tabular}
\caption{Critical points corresponding to interacting varying Chaplygin gas, Equation~(\ref{eq:M2}), for the non-gravitational interaction $Q$ given by Equation~(\ref{eq:Int3}), where $\hat{r} = \sqrt{-3b b^{2}(1+\alpha)^{2} + (n-3(1+\alpha))^{2}}$.}
  \label{tab:table8}
\end{table}

 \begin{table}
  \centering
    \begin{tabular}{ | c | c |  c |  c |  c |  c |  c |  c | p{2cm} |}
    \hline
 $n$ & $\alpha$ & $b$ \\
       \hline
 $n=0$  & $0\leq \alpha \leq 1$ & $0<b<\frac{18 \alpha ^2+36 \alpha +18}{45 \alpha ^2+90 \alpha +45}$ \\
          \hline
$0<n<\frac{3}{2}$  &  $0\leq \alpha \leq 1$ &  $\frac{-n^2+3 \alpha  n+3 n}{9 \alpha ^2+18 \alpha +n^2+9}<b\leq \frac{6 \alpha -2 n+6}{15 \alpha +15}$ \\
    \hline
$\frac{3}{2}\leq n<3$  &  $\frac{1}{3} (2 n-3)<\alpha \leq 1$ &  $ \frac{-n^2+3 \alpha  n+3 n}{9 \alpha ^2+18 \alpha +n^2+9}<b\leq \frac{6 \alpha -2 n+6}{15 \alpha +15}$ \\
    \hline
     \end{tabular}
\caption{Constraints on the model parameters for the interacting varying Chaplygin gas, Equation (\ref{eq:M2}), with the non-gravitational interaction term $Q$ given by Equation (\ref{eq:Int3}). $E.6.4$
 describes the state of the universe where the varying Chaplygin gas, Equation (\ref{eq:M2}), is a phantom dark energy.}
  \label{tab:table9}
\end{table}

The $4$ critical points presented in Table \ref{tab:table8}  describe the model with interacting varying Chaplygin gas, Equation (\ref{eq:M2}), when the non-gravitational interaction is given by Equation (\ref{eq:Int3}). The study shows that only the critical point $E.6.4$ is late time attractor describing the state of the universe with
\begin{equation}
q=-1 + \frac{n}{2(1+\alpha)},
\end{equation}

\begin{equation}
\omega_{eff} = -1 + \frac{n}{3(1+\alpha)},
\end{equation}

\begin{equation}
\omega_{de} = \frac{2b(n-3(1+\alpha))}{n + 3(-1 + 2b) (1+\alpha) + \hat{r}}.
\end{equation}

Moreover, it is a scaling attractor and the solution of the cosmological coincidence problem reads as

\begin{equation}
r = -\frac{1}{2} - \frac{\hat{r}}{2 (n - 3(-1 + 2b) (1+\alpha))}, 
\end{equation}
where $\hat{r} = \sqrt{-3b b^{2}(1+\alpha)^{2} + (n-3(1+\alpha))^{2}}$. Similar to the other two models considered in this subsection, $E.6.4$ solution can describe the state of the universe where interacting varying Chaplygin gas, Equation (\ref{eq:M2}), is either a phantom dark energy or a quintessence dark energy. In particular, the constraints on the model parameters presented in Table \ref{tab:table9}  describe the universe where the varying Chaplygin gas, Equation (\ref{eq:M2}), is a phantom dark energy. It should be mentioned that $E.6.3$ is physically not reasonable solution, while $E.6.2$ is a stable focus and does not describe the accelerated expansion.

Definitely the phase space analysis reveals that the models are very interesting. However, as it is mentioned earlier we have to impose some empirical constraints on the free parameters in order to reduce the phase space size. Definitely it simplifies the discussion. However, we should wonder whether or not we assumed reasonable constraints for the free parameters. In order to understand this, in the second part of this paper, we will apply Bayesian Machine Learning to learn the constraints. The detailed analysis is presented in the next section. The general philosophy behind the approach can be found in ~\cite{Elizalde:2020pps, Elizalde:2020mfs, Elizalde:2021kmo, Aljaf:2020nsl}. Following earlier works we also use PyMC3 to perform the analysis and learning process.

\section{Constraints from Bayesian Machine Learning}\label{sec:FFS}

In the first part of this paper we discussed the phase space analysis of the models. It appears possible to find all critical points analytically allowing to discuss conditional critical points and attractors. The last means that for imposed constraints some of the obtained solutions can serve as attractors and have specific type of stability, which can be changed when another set of constraints on the model free parameters will be imposed. A necessity somehow to constrain the possible parameter space and decrease the uncertainty we applied the Bayesian Machine Learning. We will omit all technical details behind this approach referring the readers to several recent papers \cite{Elizalde:2020pps, Elizalde:2020mfs, Elizalde:2021kmo, Aljaf:2020nsl}) for more details. The crucial aspect of this approach to be mentioned is the option connecting posterior and prior without a need to evaluate the likelihood. This allows us to depart from the real observational data concept and use simulated one. It is obvious that the method is good for forecasting purposes too. However, the validation of the learned result, as in the case of any other Machine Learning approach is required. In this early stage of analysis we used the background dynamics of each model to generate only the expansion rate data used for the learning process. Therefore available expansion rate data presented in Table \ref{tab:table10} has been used during the result validating process. It should be mentioned that our interest to the expansion rate data is due to the estimation of $H(z)$ from the cosmic chronometers. Since they are model independent estimations then there is a possibility to learn features of the models not depending on the features of the other models used to extract $H(z)$ data. However, the analogues of other observations can be crafted/generated and used to learn the constraints, too. This and other interesting questions for this moment have not been studied. They have been left to be tackled in forthcoming papers. In the next two subsections we will discuss learned constraints and see whether or not the models can be used to solve the $H_{0}$ problem (see \cite{Aghanim:2018eyx, Riess:2018byc, Wong:2019kwg, Freedman:2019jwv} for more details).
\begin{table}[t]
  \centering
    \begin{tabular}{ | c | c |  c |  c |  c |  c |  c |  c | p{2cm} |}
    \hline
$z$ & $H(z)$ & $\sigma_{H}$ & $z$ & $H(z)$ & $\sigma_{H}$ \\
      \hline
$0.070$ & $69$ & $19.6$ & $0.4783$ & $80.9$ & $9$ \\
         
$0.090$ & $69$ & $12$ & $0.480$ & $97$ & $62$ \\
    
$0.120$ & $68.6$ & $26.2$ &  $0.593$ & $104$ & $13$  \\
 
$0.170$ & $83$ & $8$ & $0.680$ & $92$ & $8$  \\
      
$0.179$ & $75$ & $4$ &  $0.781$ & $105$ & $12$ \\
       
$0.199$ & $75$ & $5$ &  $0.875$ & $125$ & $17$ \\
     
$0.200$ & $72.9$ & $29.6$ &  $0.880$ & $90$ & $40$ \\
     
$0.270$ & $77$ & $14$ &  $0.900$ & $117$ & $23$ \\
       
$0.280$ & $88.8$ & $36.6$ &  $1.037$ & $154$ & $20$ \\
      
$0.352$ & $83$ & $14$ & $1.300$ & $168$ & $17$ \\
       
$0.3802$ & $83$ & $13.5$ &  $1.363$ & $160$ & $33.6$ \\
      
$0.400$ & $95$ & $17$ & $1.4307$ & $177$ & $18$ \\

$0.4004$ & $77$ & $10.2$ & $1.530$ & $140$ & $14$ \\
     
$0.4247$ & $87.1$ & $11.1$ & $1.750$ & $202$ & $40$ \\
     
$0.44497$ & $92.8$ & $12.9$ & $1.965$ & $186.5$ & $50.4$ \\

$$ & $$ & $$ & $$ & $$ & $$\\ 

$0.24$ & $79.69$ & $2.65$ & $0.60$ & $87.9$ & $6.1$ \\
$0.35$ & $84.4$ & $7$ &  $0.73$ & $97.3$ & $7.0$ \\
$0.43$ & $86.45$ & $3.68$ &  $2.30$ & $224$ & $8$ \\
$0.44$ & $82.6$ & $7.8$ &  $2.34$ & $222$ & $7$ \\
$0.57$ & $92.4$ & $4.5$ &  $2.36$ & $226$ & $8$ \\ 
          \hline
    \end{tabular}
    \vspace{5mm}
\caption{$H(z)$ and its uncertainty $\sigma_{H}$ are in the units of km s$^{-1}$ Mpc$^{-1}$. The upper panel consists of thirty samples deduced from the differential age method. The lower panel corresponds to ten samples obtained from the radial BAO method. The table is according to \cite{Cai:2019bdh} %~\cite{ModifStart_3}
~(see also references therein for details).}
  \label{tab:table10}
\end{table}

\subsection{Varying Chaplygin gas $P_{de} = A\rho_{de} - \frac{BH^{-n}}{\rho_{de}^{\alpha}}$}

\begin{figure}[h!]
 \begin{center}$
 \begin{array}{cccc}
 \includegraphics[width=150 mm]{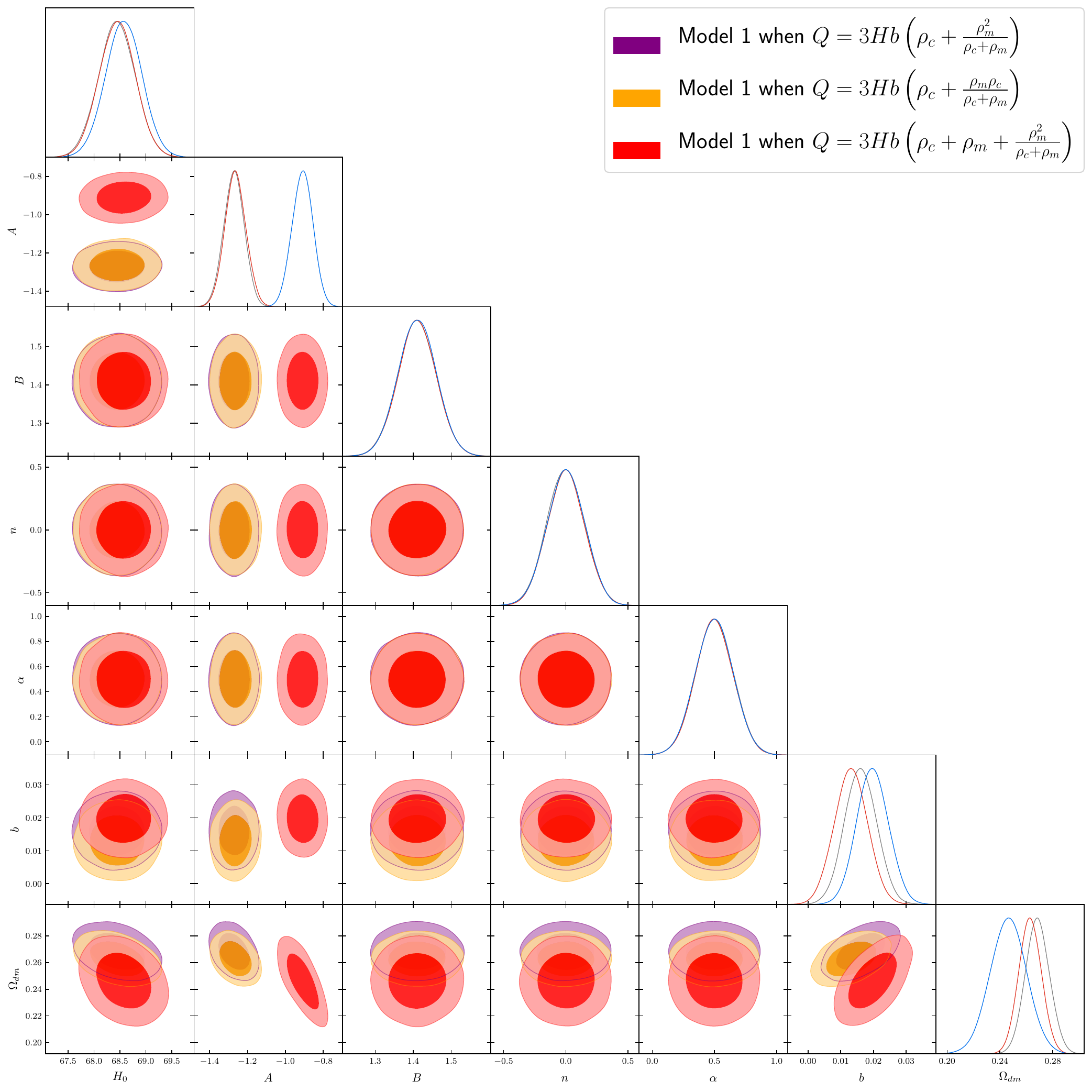} 
 \end{array}$
 \end{center}
\caption{Contour maps of the model given by Equation (\ref{eq:M1}) for $z \in [0,2.5]$. The best fit values of the model parameters have been found and presented in Table \ref{tab:table11}. In all three cases, $H_{0} \in [64.0, 80.0]$,  $A \in [-2.5, 0.5]$, $B \in [-1.0, 2.5]$, $n \in [-2.5, 2.5]$, $\alpha \in [-0.5,1.5]$, $b \in [-0.3,0.3]$ and $\Omega_{dm} \in [0.1, 0.45]$  flat priors have been imposed during the generative process used to generate the “observational” data. The analysis is based on 10 chains and, in each chain, 10,000 “observational” data-sets from the model have been generated.}
 \label{fig:FigObs_1}
\end{figure}

We start with the case where the dark energy model is given by Equation (\ref{eq:M1}). We took into account Equations (\ref{eq:fridman}),  (\ref{eq:firstfluid}) and (\ref{eq:secondfluid}) to construct the background dynamics used in the generative process. We considered three forms of non-gravitational interaction given by Equations (\ref{eq:Int1}),  (\ref{eq:Int2}) and (\ref{eq:Int3}) respectively to complete the field equations. Assuming that we have cold dark matter with $P_{dm} = 0$ we learned the constraints on $7$ free parameters in each case using generated expansion rate data. The learned constrained can be found in Table \ref{tab:table11}. During the learning process we imposed $H_{0} \in [64.0, 80.0]$,  $A \in [-2.5, 0.5]$, $B \in [-1.0, 2.5]$, $n \in [-2.5, 2.5]$, $\alpha \in [-0.5,1.5]$, $b \in [-0.3,0.3]$ and $\Omega_{dm} \in [0.1, 0.45]$ flat priors on free parameters.The results presented below are from the analysis based on 10 chains and in each chain, 10,000 “observational” data-sets from the models have been simulated. After a very detailed analysis of our results we found that:

\begin{itemize}

\item When the interaction is given by Equation.~(\ref{eq:Int1}) the learned best fit values for the model free parameters with $1\sigma$ error are: $H_{0} = 68.44 \pm 0.36$ km s$^{-1}$ Mpc$^{-1}$, $A = -1.27 \pm 0.05$, $B = 1.41 \pm 0.05$, $n = -0.0015 \pm 0.15$,  $\alpha = 0.5 \pm 0.15$, $b = 0.016 \pm 0.005$  and $\Omega_{dm}= 0.268 \pm 0.009$. The contour map is given in Figure (\ref{fig:FigObs_1}), in purple colour.

\item  On the other hand, the interaction is given by Equation~(\ref{eq:Int2}) the learned best fit values for the model free parameters with $1\sigma$ error are: $H_{0} = 68.45 \pm 0.35$  km s$^{-1}$ Mpc$^{-1}$, $A = -1.26 \pm 0.06$, $B = 1.41 \pm 0.05$, $n = 0.00037 \pm 0.15$, $ \alpha = 0.5 \pm 0.15$, $b = 0.013 \pm 0.005$  and $\Omega_{dm}= 0.263 \pm 0.008$. The contour map is given in Figure~(\ref{fig:FigObs_1}), in orange colour.

\item Finally, when the interaction is given by Equation~(\ref{eq:Int3}) the learned best fit values for the model free parameters with $1\sigma$ error are: $H_{0}= 68.57 \pm 0.35$ km s$^{-1}$ Mpc$^{-1}$, $A=-0.91 \pm 0.06$, $B = 1.41 \pm 0.05$, $n = 0.00098 \pm 0.15$, $\alpha = 0.5 \pm 0.15$, $b = 0.02 \pm 0.005$  and $\Omega_{dm}=0.247 \pm 0.014$. The contour map is given in Figure~(\ref{fig:FigObs_1}), in red colour.

\end{itemize}

As we see the models cannot solve the $H_{0}$ tension problem. Moreover, we found a hint that the model with interaction given by Equation (\ref{eq:Int3}) should be ruled out due to very low $\Omega_{dm}$. Moreover, all of them are in huge tension with high redshift expansion rate data and non of them can explain the BOSS experiment results \cite{Alam:2016hwk}. Another interesting result to be mentioned is about the learned constraints for free parameter n. We see that the learned best fit value of it can be negative for the model with Equation (\ref{eq:Int1}) interaction term. In general, during our analysis we learned that the expansion rate data will not put tight constraints on it. On the other hand, this gives a hint that considered constraints on the free parameters during the phase space analysis just allows partially cover the phase space of the models. However, we obtained all critical points analytically, therefore with future improved new tight constraints on the free parameters we can explore phase space regions more carefully. To summarize the results of the learning process we can say that there is a hint that all three models should be ruled out. One thing is clear that still additional analysis is needed for the final conclusion.

\begin{table}
  \centering
    \begin{tabular}{ | c | c | c | c | c | c |  c | c | c | c | c | p{2cm} |}
    \hline
    
 $H_{0}$ & $A$ & $B$ & $n$ & $\alpha$ & $b$ \\
      \hline
 
$68.44 \pm 0.36$ & $-1.27 \pm 0.05$ & $1.41 \pm 0.05$ &  $-0.0015 \pm 0.15$   &  $0.5 \pm 0.15$   &  $0.016 \pm 0.005$ \\
          \hline

$68.45 \pm 0.35$ & $-1.26 \pm 0.06$ & $1.41 \pm 0.05$ &  $0.00037 \pm 0.15$  &  $0.5 \pm 0.15$   &  $0.013 \pm 0.005$ \\
          \hline
          
$68.57 \pm 0.35$ & $-0.91 \pm 0.06$ & $1.41 \pm 0.05$ &  $0.00098 \pm 0.15$ & $0.5 \pm 0.15$   &  $0.02 \pm 0.005$ \\

           \hline
 
     \end{tabular}
\caption{Best fit values and $1\sigma$ errors estimated for the Model given by Equation (\ref{eq:M1}), when$z \in [0,2.5]$. The results have been obtained from a Bayesian Machine Learning approach, where the interaction $Q$ is given by Equation (\ref{eq:Int1}) (first row), Equation (\ref{eq:Int2}) (second row) and Equation (\ref{eq:Int3}) (third row), respectively. Moreover, when the interaction is given by Equation (\ref{eq:Int1}) we obtained $\Omega_{dm}= 0.268 \pm 0.009$, while when the interaction is given by Equation  (\ref{eq:Int2}) $\Omega_{dm } = 0.263 \pm 0.008$ has been obtained. Finally, $\Omega_{dm}= 0.247 \pm 0.014$ has been obtained when the interaction has been given by Equation (\ref{eq:Int3}). $H_{0} \in [64.0, 80.0]$,  $A \in [-2.5, 0.5]$, $B \in [-1.0, 2.5]$, $n \in [-2.5, 2.5]$, $\alpha \in [-0.5,1.5]$, $b \in [-0.3,0.3]$ and $\Omega_{dm} \in [0.1, 0.45]$ flat priors. The analysis is based on 10 chains and in each chain, 10,000 “observational” data-sets from the models have been simulated. $H_{0}$ and its uncertainty $\sigma$ are in the units of km s$^{-1}$ Mpc$^{-1}$.}
  \label{tab:table11}
\end{table}

\subsection{Varying Chaplygin gas $P_{c} = A\rho_{c} - \frac{B a^{-n}}{\rho_{c}^{\alpha}}$}

\begin{figure}[h!]
 \begin{center}$
 \begin{array}{cccc}
 \includegraphics[width=150 mm]{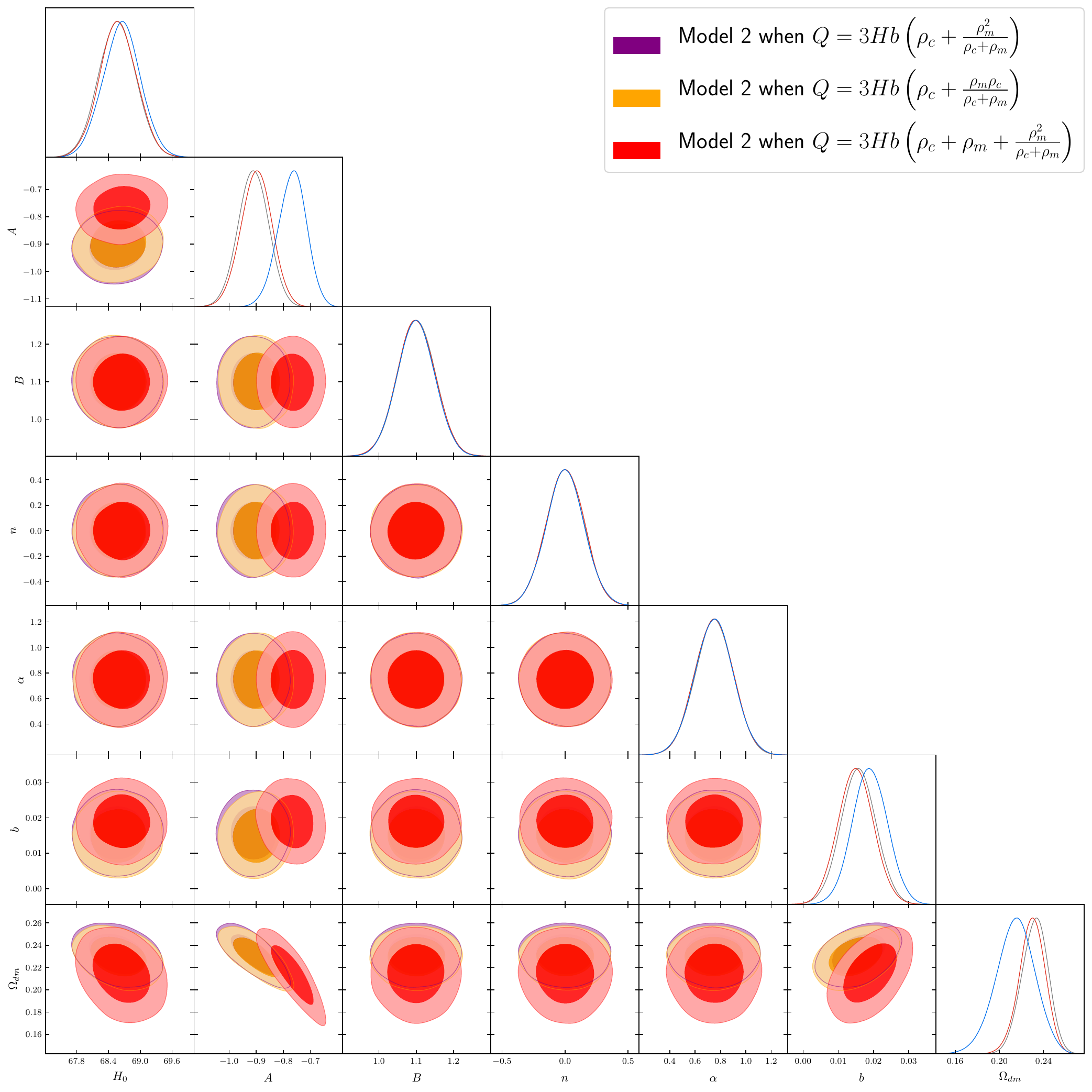}
 \end{array}$
 \end{center}
\caption{Contour maps of the model given by Equation (\ref{eq:M2}) for $z \in [0,2.5]$. The best fit values of the model parameters have been found and presented in Table \ref{tab:table12}. In all three cases, $H_{0} \in [64.0, 80.0]$,  $A \in [-2.5, 0.5]$, $B \in [-1.0, 2.5]$, $n \in [-2.5, 2.5]$, $\alpha \in [-0.5,1.5]$, $b \in [-0.3,0.3]$ and $\Omega_{dm} \in [0.1, 0.45]$ flat priors have been imposed during the generative process used to generate the “observational” data. The analysis is based on 10 chains and, in each chain, 10,000 “observational” data-sets from the model have been generated.}
 \label{fig:FigObs_2}
\end{figure}

Finally, we analyzed the model where the dark energy is given by Equation (\ref{eq:M2}). Again, we took into account Equations  (\ref{eq:fridman}),  (\ref{eq:firstfluid}) and (\ref{eq:secondfluid}) to construct the background dynamics used in the generative process. In this case, also we considered three forms of non-gravitational interaction given by Equations (\ref{eq:Int1}),  (\ref{eq:Int2}) and (\ref{eq:Int3}) respectively. Assuming that we have cold dark matter with $P_{dm} = 0$ we learned the constraints on $7$ free parameters in each case using generated expansion rate data. Similar to the first case, during the learning process we imposed  $H_{0} \in [64.0, 80.0]$,  $A \in [-2.5, 0.5]$, $B \in [-1.0, 2.5]$, $n \in [-2.5, 2.5]$, $\alpha \in [-0.5,1.5]$, $b \in [-0.3,0.3]$ and $\Omega_{dm} \in [0.1, 0.45]$
 flat priors on free parameters. The results presented below are from the analysis based on 10 chains and in each chain, 10,000 “observational” data-sets from the models have been simulated. The learned constrained can be found in Table~\ref{tab:table12}  and to simplify our discussion we present them below. In particular, using Bayesian Machine Learning based on generated expansion rate data we found that:

\begin{itemize}

\item When the interaction is given by Eq.~(\ref{eq:Int1}) the learned best fit values for the model free parameters with $1\sigma$ error are: $H_{0} = 68.53 \pm 0.35$ km s$^{-1}$ Mpc$^{-1}$, $A = -0.91 \pm 0.06$, $B = 1.11 \pm 0.05$, $n = -3.7 \times 10^{-5} \pm 0.15$, $]alpha = 0.75 \pm 0.15$, $b = 0.016 \pm 0.005$ and $\Omega_{dm}= 0.233 \pm 0.012$. The contour map is given in Figure (\ref{fig:FigObs_2}), in purple colour.

\item  On the other hand, the interaction is given by Eq.~(\ref{eq:Int2}) the learned best fit values for the model free parameters with $1\sigma$ error are: 
$H_{0} = 68.58 \pm 0.34$ km s$^{-1}$ Mpc$^{-1}$, $A = -0.89 \pm 0.06$, $B = 1.09 \pm 0.05$, $n = 0.0026 \pm 0.15$, $\alpha = 0.75 \pm 0.15$, $b = 0.015 \pm 0.005$  and $\Omega_{dm}= 0.229 \pm 0.012$. The contour map is given in Figure (\ref{fig:FigObs_2}), in orange colour.

\item Finally, when the interaction is given by Eq.~(\ref{eq:Int3}) the learned best fit values for the model free parameters with $1\sigma$ error are: $H_{0} = 68.65 \pm 0.35$ km s$^{-1}$ Mpc$^{-1}$, $A = -0.77 \pm 0.05$, $B = 1.09 \pm 0.05$, $n = 0.0015 \pm 0.15$, $\alpha = 0.75 \pm 0.15$, $b = 0.019 \pm 0.005$ and $\Omega_{dm}= 0.215 \pm 0.017$. The contour map is given in Figure (\ref{fig:FigObs_2}), in red colour.

\end{itemize}

Again we see the models cannot solve the $H_{0}$ tension problem. Moreover, we found a hint that the models should be ruled out due to very low  $\Omega_{dm}$. Moreover, all of them are in huge tension with high redshift expansion rate data and non of them can explain the BOSS experiment results. In other words, the models considered in this subsection should be rejected. We would like to mention that the learning based on generated other data sets can induce a significant clarification on this issue, therefore an additional analysis is still required before the final rejection of the models. It has been left to be discussed in a forthcoming paper.

\begin{table}
  \centering
    \begin{tabular}{ | c | c | c | c | c | c |  c | c | c | c | c | p{2cm} |}
    \hline
    
 $H_{0}$ & $A$ & $B$ & $n$ & $\alpha$ & $b$ \\
      \hline
 
$68.53 \pm 0.35$ & $-0.91 \pm 0.06$ & $1.11 \pm 0.05$ &  $-3.7 \times 10^{-5} \pm 0.15$   &  $0.75 \pm 0.15$   &  $0.016 \pm 0.005$ \\
          \hline

$68.58 \pm 0.34$ & $-0.89 \pm 0.06$ & $1.09 \pm 0.05$ &  $0.0026 \pm 0.15$  &  $0.75 \pm 0.15$   &  $0.015 \pm 0.005$  \\
          \hline
          
$68.65 \pm 0.35$ & $-0.77 \pm 0.05$ & $1.09 \pm 0.05$ &  $0.0015 \pm 0.15$ & $0.75 \pm 0.15$   &  $0.019 \pm 0.005$  \\

           \hline
 
     \end{tabular}
\caption{Best fit values and $1\sigma$ errors estimated for the Model given by Equation (\ref{eq:M2}), when $z \in [0,2.5]$. The results have been obtained from a Bayesian Machine Learning approach, where the interaction $Q$ is given by Equation (\ref{eq:Int1}) (first row), Equation (\ref{eq:Int2}) (second row) and Equation (\ref{eq:Int3}) (third row), respectively. Moreover, when the interaction is given by Equation (\ref{eq:Int1}) we obtained $\Omega_{dm} = 0.233 \pm 0.012$, while when the interaction is given by Equation (\ref{eq:Int2}) $\Omega_{dm} = 0.229 \pm 0.012$ has been obtained. Finally, $\Omega_{dm} =0.215 \pm 0.017$ has been obtained when the interaction has been given by Equation (\ref{eq:Int3}). We used $H_{0} \in [64.0, 80.0]$,  $A \in [-2.5, 0.5]$, $B \in [-1.0, 2.5]$, $n \in [-2.5, 2.5]$, $\alpha \in [-0.5,1.5]$, $b \in [-0.3,0.3]$ and $\Omega_{dm} \in [0.1, 0.45]$ flat priors. The analysis is based on 10 chains and in each chain, 10,000 “observational” data-sets from the models have been simulated. H0
 and its uncertainty $\sigma$ are in the units of km s$^{-1}$ Mpc$^{-1}$.}
  \label{tab:table12}
\end{table}

\section{\large{Discussion}}\label{sec:Discussion}

In this paper, we considered six different cosmological scenarios. Three different forms of the interaction term $Q$ have been used to model non-gravitational interaction supposedly existing between dark energy and cold dark matter. The Chaplyging gas (and its various modifications) is still among very actively considered dark energy gas/fluid models. Actually two modifications of it has been considered. In particular, with specific phenomenological assumptions about its free parameter it is possible to construct new interesting modifications. Following to this approach, in this work we considered two models where two parametrizations of $B$ are taken into account. We performed the phase space analysis of all models. In all cases, late time scaling attractors have been found analytically and by imposing empirical constraints their nature have been explored. In particular, it has been found that for the first type of cosmological models the late time scaling attractors describe the state of the universe where varying Chaplygin gas has only a phantom nature. On the other hand, the study of the second class of models shows that for some values of the model free parameters the varying Chaplygin gas will be a phantom dark energy, while for some cases it will be a quintessence dark energy. An interesting result that has not been observed during the study of the first model. Moreover, in the second part of the paper we applied Bayesian Machine Learning approach to learn the constraints on the model parameters. It is a good way to understand the models and extract additional information justifying the reason specific phenomenological modifications should exist. On the other hand, it appears to be very useful for future studies pointing out the shape of the parameter space. In our analysis the background dynamics of each model has been used to generate the expansion rate data to be used in the learning process. We have a very brief discussion on this, referring the reader to a series of papers demonstrating how the Bayesian Machine Learning can be used to achieve very interesting and unique results. The results from the Bayesian Machine Learning based on generated expansion rate data revealed interesting results. In particular, the results of our analysis showed that it is very hard to say the models eventually should be rejected or not. However, we found a hint that they should be rejected since (1) they cannot solve the $H_{0}$ tension problem, and (2) there is a huge tension between learned and observational results at high redshifts. However, we need to take into account that the learned constraints indicate—the expansion rate data is not good to learn the constraints on the model free parameters. It can be that the learning process based on other generated data sets can do the job better and eventually it would be possible to learn tight constraints on the parameters. This among other problems has been left to be tackled in forthcoming papers. In this stage of the analysis the Bayesian Machine Learning indicates several interesting aspects of considered varying Chaplygin gas models. One of them is related to the forms of non-gravitational interaction. In particular, a recent analysis of  \cite{Elizalde:2021kmo} demonstrated that a deviation form cold dark matter paradigms can easily be confused with non-gravitational interaction between dark energy and dark matter. In this regard, we need to mention that a similar situation could be observed here too. In this case, if it is true, then we can significantly reduce the phenomenology and craft new cosmological models with varying Chaplygin gas, where the model rejection question will be more transparent and easily understandable. Our initial attempts to study the Chaplygin gas with Machine Learning approaches are very promising and should be extended, revealing whether or not it can be really used to unify dark energy and dark matter. Other interesting questions related to this issue we left to be tackled in the future papers. In general with future research we still need to understand whether or not the expansion rate data can capture the features of non-linear non-gravitational interaction between dark energy and dark matter properly. In addition, if it cannot, then how to be with the bias that can be introduced in the analysis where the expansion rate data has been used with other observational data sets and the features of the non-gravitational interaction have been captured?

In summary, the Bayesian Machine Learning reveals that something is not smooth with varying Chaplygin gas models we considered here. However, a final conclusion requires additional work which can provide a fresh look at (varying) Chaplygin gas cosmology and significantly reduce so far existing phenomenology.

\section*{Acknowledgement}

This work was supported by the Ministry of Education and Science of Kazakhstan under grants AP09261147.

%\bibliographystyle{elsarticle-num-ID}
%\bibliography{lit.bib}

\end{document}